\documentclass[11pt,aps,prl,reprint]{revtex4-1}
\usepackage{amsmath}    
\usepackage{graphicx}   
\usepackage{bm}

\begin{document}
\title{Dissociative recombination of cold HeH$^+$ ions}
\author{Roman \v{C}ur\'{\i}k} \email{roman.curik@jh-inst.cas.cz}
\affiliation{J. Heyrovsk\'{y} Institute of Physical Chemistry, ASCR,
Dolej\v{s}kova 3, 18223 Prague, Czech Republic}
\author{D\'{a}vid Hvizdo\v{s}}
\affiliation{J. Heyrovsk\'{y} Institute of Physical Chemistry, ASCR,
Dolej\v{s}kova 3, 18223 Prague, Czech Republic}
\affiliation{Institute of Theoretical Physics, Faculty of Mathematics and Physics, Charles University in Prague, V Hole\v{s}vi\v{c}k\'{a}ch 2, 180 00 Prague, Czech Republic}
\author{Chris H.~Greene}
\affiliation{Department of Physics and Astronomy, Purdue University, West Lafayette,
Indiana 47907, USA}
\affiliation{Purdue Quantum Science and Engineering Institute, Purdue University, West Lafayette, Indiana 47907, USA}
\date{\today}

\begin{abstract}
\vspace{-4mm}
The HeH$^+$ cation is the simplest molecular prototype of the indirect dissociative recombination
(DR) process that proceeds through electron capture into Rydberg states of the corresponding neutral
molecule.
This Letter develops the first application of our
recently developed energy-dependent frame transformation theory to the indirect DR
processes. The theoretical model is based on the multichannel quantum-defect theory with the vibrational basis
states computed using exterior complex scaling (ECS) of the nuclear Hamiltonian. The {\it ab initio} electronic
$R$-matrix
theory is adopted to compute quantum defects as functions of the collision energy 
and of the internuclear distance. The resulting DR rates are convolved over the beam energy distributions relevant to
a recent experiment at the Cryogenic Storage Ring, giving good agreement between the experiment and
the theory.
\end{abstract}

\maketitle

The HeH$^+$ ion, probably the oldest molecule in the Universe, had eluded astrophysical observation for decades.
Only very recently, \citet{Guesten_HeH_Nature_2019} finally reported detection of HeH$^+$ ions in the nebula 
NGC~7027. The authors created a simple reaction chain model in which the dominant roles are played by two competing
processes: the
association (RA) of He$^+$ and H atoms to form HeH$^+$ and the dissociative recombination which destroys the molecule:
\begin{equation}
\label{eq-dr}
\mathrm{HeH}^+ + e^- \rightarrow \mathrm{He} + \mathrm{H}\;.
\end{equation}
The observed beam brightness of the pure rotational transition $j = 1 \rightarrow 0$ was about 4$\times$ higher
than the value based on the rate constants for the RA (1.4 $\times$ 10$^{-16}$ cm$^3$/s) and the 
DR (3.0 $\times$ 10$^{-10}$ cm$^3$/s) processes at kinetic temperature of 10$^4$ K. This appears to indicate 
that either the currently
known RA rate \cite{Vranckx_LDLV_JPB_2013} underestimates the production of HeH$^+$ cations or else
the measured DR rate 
\cite{Stromholm_Larsson_HeH_1996} overestimates their destruction rate.

Initial state-specific DR rate coefficients have recently been measured at the Cryogenic Storage Ring (CSR)
\cite{Novotny_HeH_Science_2019}. The experimental collaboration observed a dramatic decrease of the 
DR rate at very low collision energies
(below 20 meV), far smaller than those observed in
previous measurements \cite{Stromholm_Larsson_HeH_1996}
at room temperature. Such a reduced destruction rate of the HeH$^+$ ions should be reflected by their higher
abundance in cold interstellar environments. 

A number of computational methods have been previously applied to study the DR of HeH$^+$. The pioneering
work of \citet{Guberman_HeH_1994} was based on the traditional Born-Oppenheimer framework, with the neutral
HeH curves coupled to each other by non-adiabatic
coupling terms and with rotational effects neglected. All the treatments that followed
\cite{Takagi_HeH_2004,Takagi_Tashiro_2015,Haxton_Greene_HeH_2009,Curik_Greene_JCP_2017} were based on the 
rovibrational frame transformation combined with multichannel quantum defect theory (MQDT). They differed in
their treatments of the nuclear dynamics and in the accuracy of the quantum defect matrix $ \bm{\mu}(R)$ --
a single electronic-structure quantity necessary to carry out these calculations.
The most recent calculations by \citet{Curik_Greene_JCP_2017} exhibit decent agreement with the hot ions
experiment of \citet{Stromholm_Larsson_HeH_1996} but the predicted value of
the DR rate from the initial $j'=0$ state at low electron temperatures is about one order of magnitude higher than
the values deduced in
the recent cryogenic experiment \cite{Novotny_HeH_Science_2019}.

This Letter introduces a practical application of the reformulated energy-dependent frame transformation theory
we have developed \cite{Hvizdos_CG_PRA_2019} for treating vibrational excitation and dissociative recombination processes.
The method was derived by use of a simple 2D-model \cite{Curik_HG_PRA_2018}
in which the total, electronic and nuclear Hamiltonian
was tailored to approximately
describe the $^1\Sigma$ ungerade states of H$_2$. Such a model system can be solved exactly in two
dimensions (one electronic and one nuclear coordinate), altogether avoiding the Born-Oppenheimer approximation. Therefore, it can provide
an exact benchmark for an approximate theory, such as the frame transformation approach. All the technical details
of the method are present in our companion paper \cite{Hvizdos_CG_PRA_2019}, while this Letter deals only with 
the procedural steps relevant for treating electron collisions with HeH$^+$. 
\begin{figure}[tbh]
\includegraphics[width=0.48\textwidth]{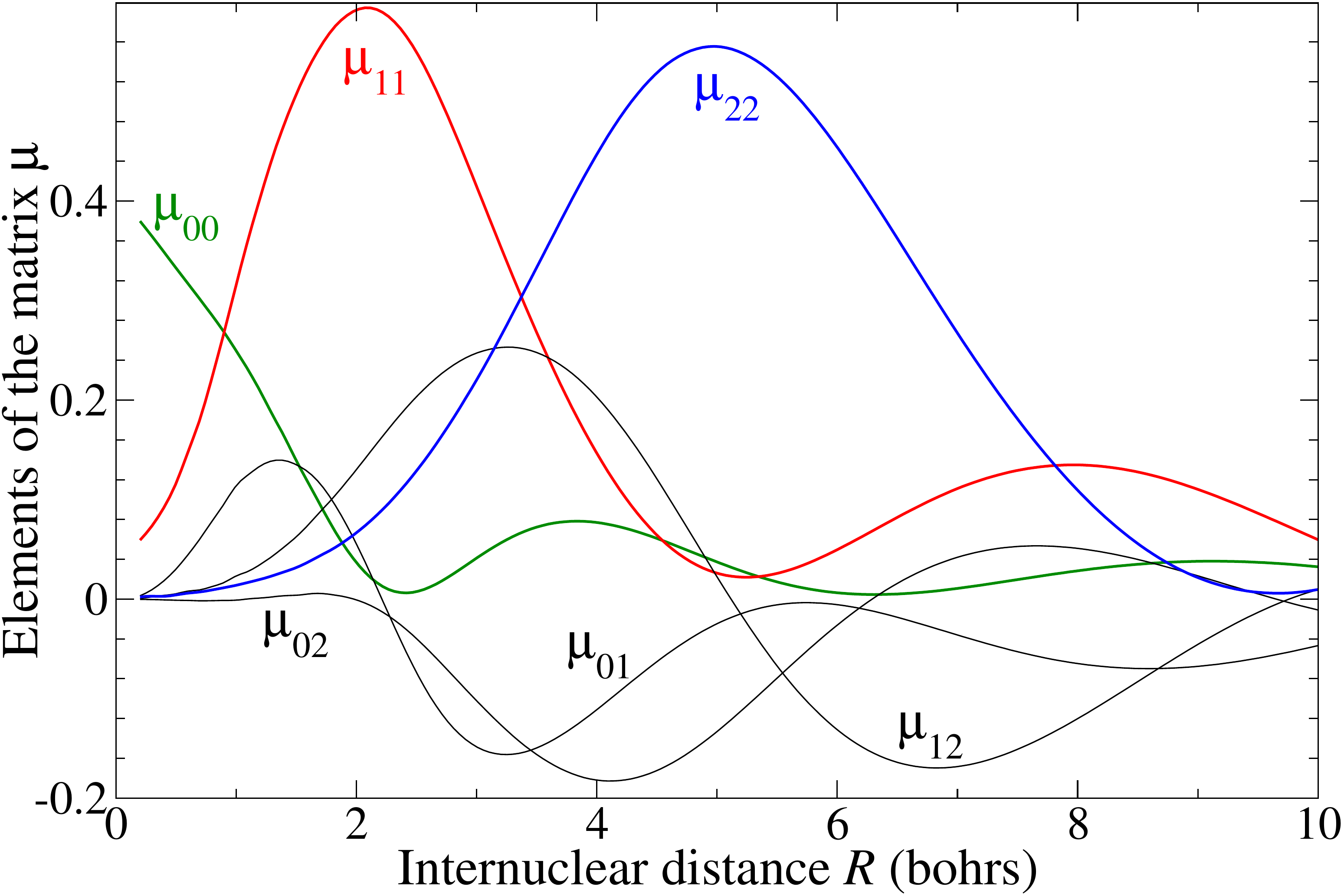}
\caption{\label{fig-mue_lam0a}
(color online). Matrix elements of the $\bm{\mu}(R)$ matrix for $\Lambda = 0$.
}
\end{figure}

In contrast to the previous DR studies of HeH$^+$, the present treatment is based on the energy-dependent quantum defect matrix,
approximated throughout this study with a linear energy dependence, as
\begin{equation}
\label{eq-qde}
\bm{\mu}^{\Lambda}(R,\epsilon) = \bm{\mu}^{\Lambda}(R) + \epsilon \bm{\mu'}^{\Lambda}(R)\;.
\end{equation}
Here $\epsilon$ represents the body-frame electron energy, and $\Lambda$ denotes the projection of the Rydberg
electron angular momentum $l$ onto the molecular axis.

The quantum defect matrices have been computed using the diatomic UK R-matrix package 
\cite{Morgan_Chen_Rm_1997} with the $R$-matrix boundary set at $r_0$ = 20 bohrs. Bound electrons
are described in the Slater-type basis \cite{Ema_Paldus_STO_2003} (STO) of triple-zeta quality 
(denoted as VB2 in Ref.~\cite{Ema_Paldus_STO_2003}). 
The angular space of the colliding (or Rydberg) electron is limited by $l_{max} = 2$
(with $\Lambda_{max}=2$), which we previously found \cite{Curik_Greene_JCP_2017}
to be sufficient when working in the center-of-charge frame of reference.

The zero-energy quantum defect matrix
$\bm{\mu}^{\Lambda}(R)$ and the linear coefficient $\bm{\mu'}^{\Lambda}(R)$
in Eq.~(\ref{eq-qde}) are obtained by carrying out fixed-nuclei scattering calculations for the
$e^- + \mathrm{HeH}^+$ system, for collision energies $\epsilon_1$ = 20 meV and $\epsilon_2$ = 420 meV, 
followed by numerical differentiation. The stability of this procedure was checked by changing
$\epsilon_2$ to 220 and 620 meV. Upon these changes the matrix $\bm{\mu'}^{\Lambda}(R)$ varied only
within 1-2\%, while variations of $\bm{\mu}^{\Lambda}(R)$ were smaller than 0.2\%.

\begin{figure}[tbh]
\includegraphics[width=0.48\textwidth]{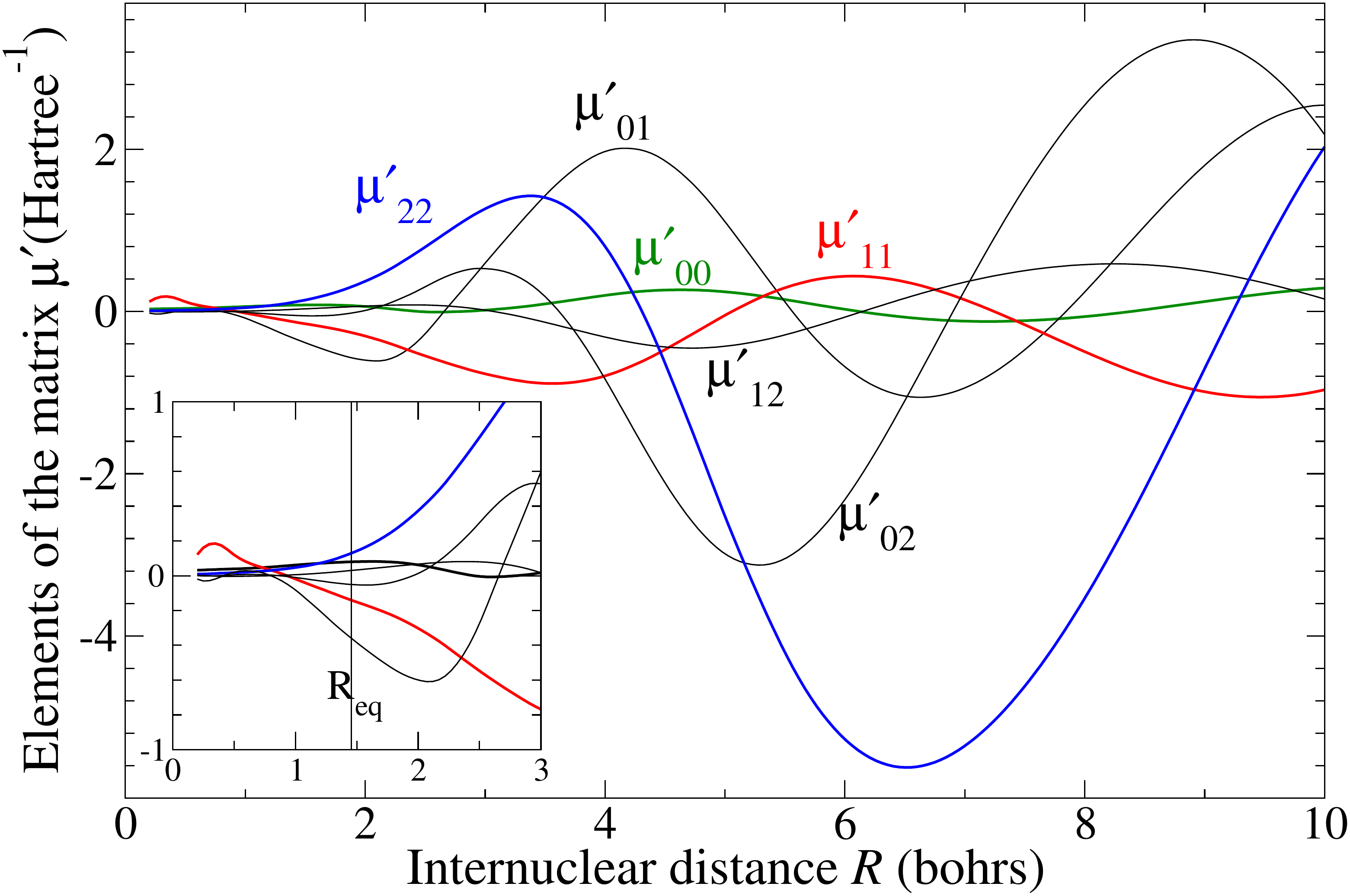}
\caption{\label{fig-mue_lam0b}
(color online). Matrix elements of the $\bm{\mu'}(R)$ matrix for $\Lambda = 0$.
}
\end{figure}

The resulting $R$-dependences of the 
$\bm{\mu}^{\Lambda}(R)$ and $\bm{\mu'}^{\Lambda}(R)$ elements are shown in Figs.~\ref{fig-mue_lam0a}
and \ref{fig-mue_lam0b} for $\Lambda = 0$. While the data shown in Fig.~\ref{fig-mue_lam0a} are very similar
to the quantum defects published previously \cite{Curik_Greene_JCP_2017}, the linear energy slope
coefficients displayed in Fig.~\ref{fig-mue_lam0b} are new. Quantum defects (and their energy dependence) for
$\Lambda = 1, 2$ have also been computed and used in the present work. 
However, since their impact on the final DR rates is minor, they are not shown here.

The rovibrational nuclear basis, serving as channel functions in the MQDT framework, is generated
by numerically solving the nuclear Schr\"{o}dinger equation
\begin{equation}
\label{eq-Schro-nuc}
\left[ - \frac{\mathrm{d}^2}{\mathrm{d}Z^2} 
+ 2 M U^+(Z) + \frac{j(j+1)}{Z^2}
- K^2_{\nu j} \right] \phi_{\nu j}(Z) = 0 \;,
\end{equation}
where $U^+(Z)$ is the ground-state $^1\Sigma_g^+$ potential curve of the target cation and 
the atom-ion reduced mass $M = 1467.28$~a.u. was taken from Ref.~\cite{Coxon_Hajigeorgiou_1999}.
The nuclear Schr\"{o}dinger equation (\ref{eq-Schro-nuc}) is solved on a complex contour
$Z$ of the internuclear distances, according to the exterior complex scaling (ECS) technique
\cite{Simon_ECS_PLA_1979,McCurdy_Martin_JPB_2004}:
\begin{equation}
\label{eq-ECS-countour}
Z = \left\{
\begin{array}{ll}
 R, & \mathrm{for}\; R \leq R_0, \\
 R_0 + e^{i\theta(R-R_0)}, & \mathrm{for}\; R_0 < R \leq R_m ,
\end{array}
\right.
\end{equation}
where $R$ is a real parameter along the complex contour $Z$, $R_0 = 10$ bohr denotes the bending point,
$\theta = 40^{\circ}$ is the bending angle, and $R_m = 25$ bohrs parametrizes the final point $Z_m$ of the
complex contour. Boundary conditions for the solutions are $\phi_{\nu j}(0)$ = $\phi_{\nu j}(Z_m)$ = 0.
\begin{figure}[tbh]
\includegraphics[width=0.48\textwidth]{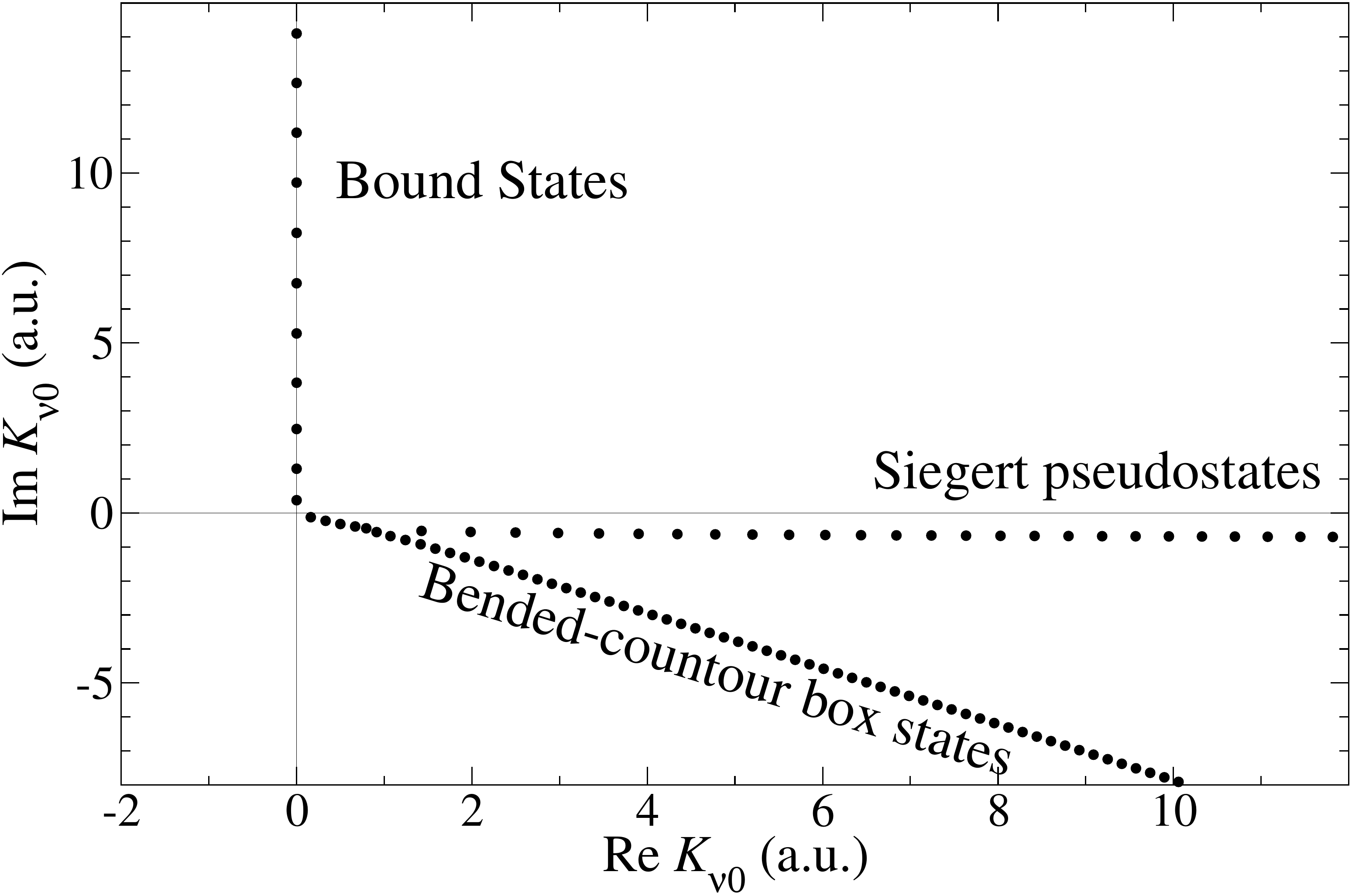}
\caption{\label{fig-ecs-poles}
Distribution of the ECS poles in the complex momentum plane for $j = 0$.
}
\end{figure}
An example of the complex eigenmomenta $K_{\nu j}$ spectrum is shown in Fig.~\ref{fig-ecs-poles} for $j=0$.
It is similar to that obtained using the Siegert pseudostate
\cite{Siegert_1939,Tolst_sieg_1998}
spectrum that has been employed in some of the previous DR studies 
\cite{Hamilton_Greene_PRL_2002,Curik_Greene_PRL_2007,Curik_Gianturco_2013}. However, in the ECS approach
some of the lowest continuum Siegert pseudostates are replaced by a branch of states 
corresponding to the box states on the rotated coordinate.
The desired completeness of the set of states $\phi_{\nu j}$ for the present calculations is
reached with 160 states (ordered by the absolute value of the corresponding energy) for each rotational quantum number $j$.

With the energy-dependent quantum defects (\ref{eq-qde}) and the rovibrational basis $\phi_{\nu j}(Z)$ capable
to cover rovibrational excitation and dissociation of the system, all requisite elements needed are now ready
to implement the
energy-dependent frame transformation procedure. This procedure has been derived in detail and benchmarked against the
exact results of the 2D-model in Ref.~\cite{Hvizdos_CG_PRA_2019}.
The resulting initial state-dependent DR cross section can be written as a sum
\begin{equation}
\label{eq-sigma-sumJ}
\sigma_{\nu'j'} (E_c) = \frac{1}{2j'+1} \sum_{\eta J l'}(2J+1) \sigma_{\nu' j' l'}^{J \eta}(E_c),
\end{equation}
where $E_c$ is the collision energy, $l'$ denotes the initial angular momentum of the colliding electron, 
$\vec{J} = \vec{j'} + \vec{l'}$ 
is the total angular momentum, and $\eta (-1)^J$ represents the parity of the whole system. The
odd $\eta$ constituents of the cross sections are negligible in this study as they depend only on the weak 
$\Lambda = 1, 2$ components of the quantum defects \cite{Chang_Fano_1972}. Nevertheless, our calculations have 
included them.
The initial state-dependent recombination rate is simply
\begin{equation}
\label{eq-DRrate}
\alpha_{\nu'j'} (E_c) = \sqrt{2 E_c}\, \sigma_{\nu'j'} (E_c)\;.
\end{equation}

The physical $S$-matrix, from which the cross sections (\ref{eq-sigma-sumJ}) are derived, 
is a result of the MQDT procedure called the elimination of closed-channels:
\begin{equation}
\label{eq-elimc}
\bm{S}^{\mathrm{phys}} = \bm{S}^{oo} - \bm{S}^{oc}\left[
\bm{S}^{cc}-e^{-2 i \bm{\beta}(E)} \right]^{-1} \bm{S}^{co} \;,
\end{equation}
where the superscripts $o$ and $c$ denote open and closed sub-blocks in the
short-range $S$-matrix, respectively.
The diagonal matrix $\bm{\beta}(E)$ describes effective Rydberg quantum
numbers with respect to the closed-channel thresholds $E_i$:
\begin{equation}
\label{eq-elimb}
\beta_{ij} = \frac{\pi}{\sqrt{2(E_i-E)}}\delta_{ij}\;.
\end{equation}

\begin{figure}[tbh]
\includegraphics[width=0.48\textwidth]{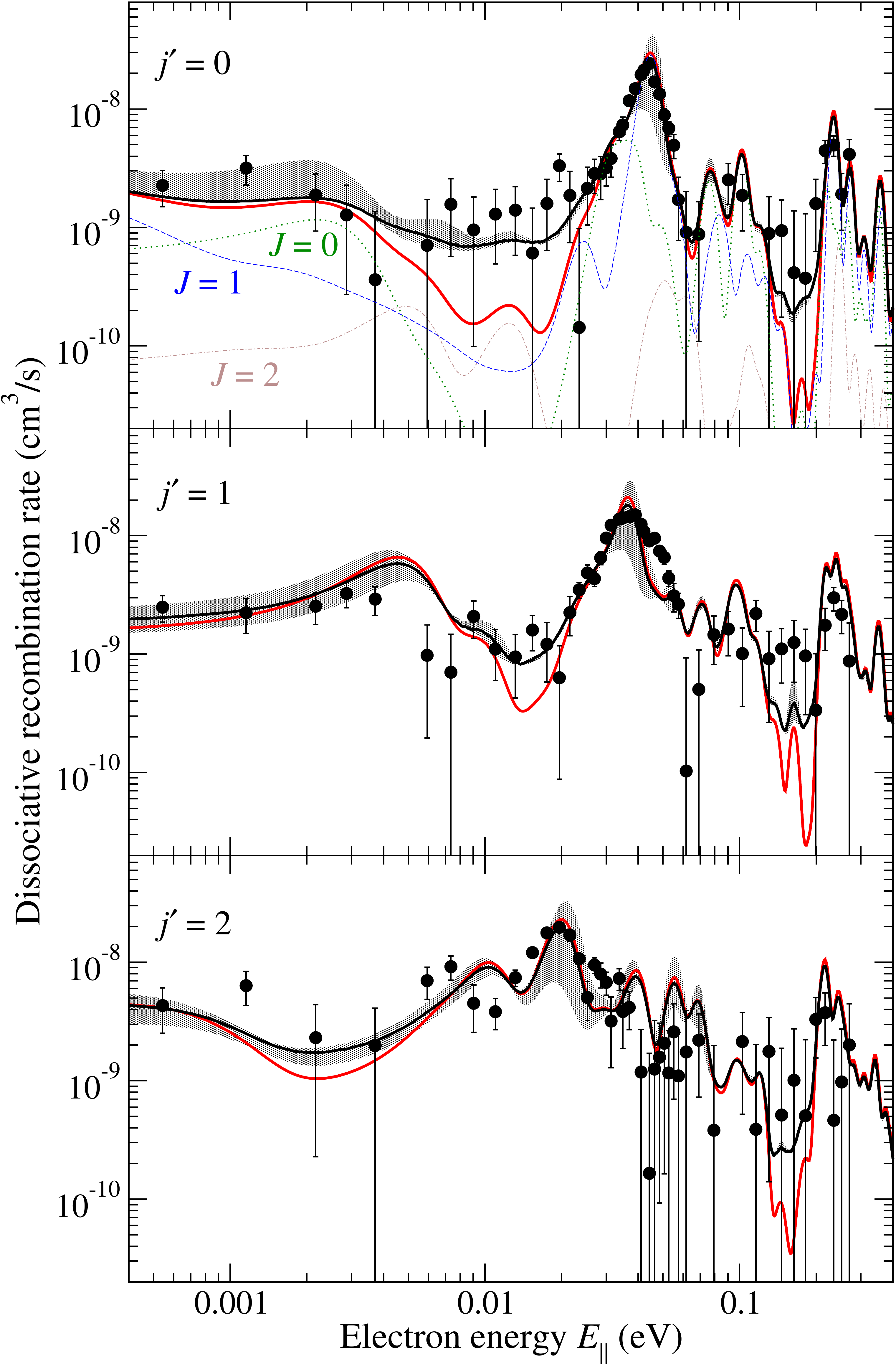}
\caption{\label{fig-drj}
(color online). Shown here is the DR rate anisotropically averaged 
with $\Delta E_{||}$ = 0.1 meV and
$\Delta E_{\perp}$ = 2 meV. The initial vibrational state is $\nu'=0$ and the initial
rotational state changes from $j'=0$ (top panel) to $j'=2$ (bottom panel).
The red curve shows the present calculations without toroidal
effects included, while the black curve accounts for the toroidal correction
\cite{Novotny_AJ_2013}.
The shaded area estimates the uncertainty of our calculations (see more details in the text).
The black circles with error bars display DR rates measured at the CSR
\cite{Novotny_HeH_Science_2019}. The top panel also shows the decomposition of the computed
DR rate into channels with different total angular momenta $J$ (dashed lines).
}
\end{figure}

The inversed term on the r.h.s. of Eq.~(\ref{eq-elimc}) generates
series of dense resonances accumulating to each of the closed-channel
ionization thresholds. These resonances are associated with the autoionizing 
and predissociating states of
the neutral HeH system. 
In order to compare our computed DR rates with the most recent experiment, and also
for likely future applications of these calculated recombination rates, our raw numerical
data must be convolved over the electron energy distributions relevant to any appropriate
environment.
In astrophysical applications, a Maxwell-Boltzmann distribution of electrons is often assumed
\cite{Andersen_Bolko_PRA_1990,Lepp_SD_JPB_2002}. 
In storage-ring experiments, on the other hand, the electron beam exhibits an anisotropic distribution.
The velocity parallel to the ion beam is usually well defined with a small spread
in the parallel energy ($\Delta E_{||}$ = 0.1 meV in Ref.~\cite{Novotny_HeH_Science_2019}).
Divergence of the electron beam is measured by the perpendicular spread, 
$\Delta E_{\perp}$ = 2 meV in Ref.~\cite{Novotny_HeH_Science_2019}.
The details and formulae for the thermal (Maxwell) and anisotropic convolutions 
can be found in Refs.~\cite{Kokoouline_Greene_PRA_2003, Curik_Greene_MP_2007}.

Fig.~\ref{fig-drj} summarizes our results, along with the recent measurements at CSR
\cite{Novotny_HeH_Science_2019}, for different initial rotational states $j'$ (the 
initial vibrational state is always $\nu'=0$).
The red curve represents the calculated DR rate convolved over the anisotropic electron beam 
distributions. The black curve shows the DR rate with the toroidal correction 
\cite{Slava_Greene_PRA_2005,Novotny_AJ_2013}
applied to the computed data. The toroidal correction accounts for the collisional events
that happen in the bending areas where the two beams merge or diverge. In these areas the relative
collisional energy is higher. Such events effectively increase the energy spread,
well beyond the values of $\Delta E_{||}$ and $\Delta E_{\perp}$, for a fraction of the DR events.
The black circles with error bars denote the data measured
recently at the CSR~\cite{Novotny_HeH_Science_2019}. 

The present calculations clearly confirm the experimentally observed \cite{Novotny_HeH_Science_2019}
low-energy behavior of the DR rate for the cold HeH$^+$ ions. Introduction of energy-dependent
body-frame
quantum defects into the theory has shifted some low-energy closed-channel resonances,
resulting in low energy DR rates that differ from the results of previous theoretical studies,
and this appears to produce improved agreement with experiment.
In particular, the steep increase of the DR rate for $j'=0$ at zero energy (Fig.~13 in
Ref.~\cite{Curik_Greene_JCP_2017}) was caused by a $l=1$ closed-channel resonance positioned at 1 meV.
The energy dependence of $\bm{\mu}(R)$ shown in Fig.~\ref{fig-mue_lam0b} causes this resonance to
move to the negative collision energies, making it a bound Rydberg state of the neutral HeH
that no longer affects the computed DR rate.

We have also attempted to estimate the sensitivity
of the computed rates to the accuracy of the electronic structure calculations. To this extent
a random noise matrix $\Delta \bm{\mu}$ of the maximum value of $\delta$ = 0.002 was added to
and subtracted from the original $\bm{\mu}^{\Lambda}(R,\epsilon)$ in Eq.~(\ref{eq-qde}).
The value of $\delta$ reflects our observations of variations of $\bm{\mu}^{\Lambda}(R,\epsilon)$
due to the electronic basis set size and due to the extent of the space of configuration interaction.
The resulting variations of the computed DR rates are shown as a shaded area in Fig.~\ref{fig-drj}.
One observes that the low-energy region is more sensitive to the accuracy of the quantum defects.
We observe 30--50\% variations of the DR rate below 1 meV and changes by a multiplicative 
factor of 2--3 in the area of the peak (20--45 meV, depending on the $j'$). At higher energies
these variations appear to be smeared out by the broader effective electron energy distribution.

\begin{figure}[tbh]
\includegraphics[width=0.48\textwidth]{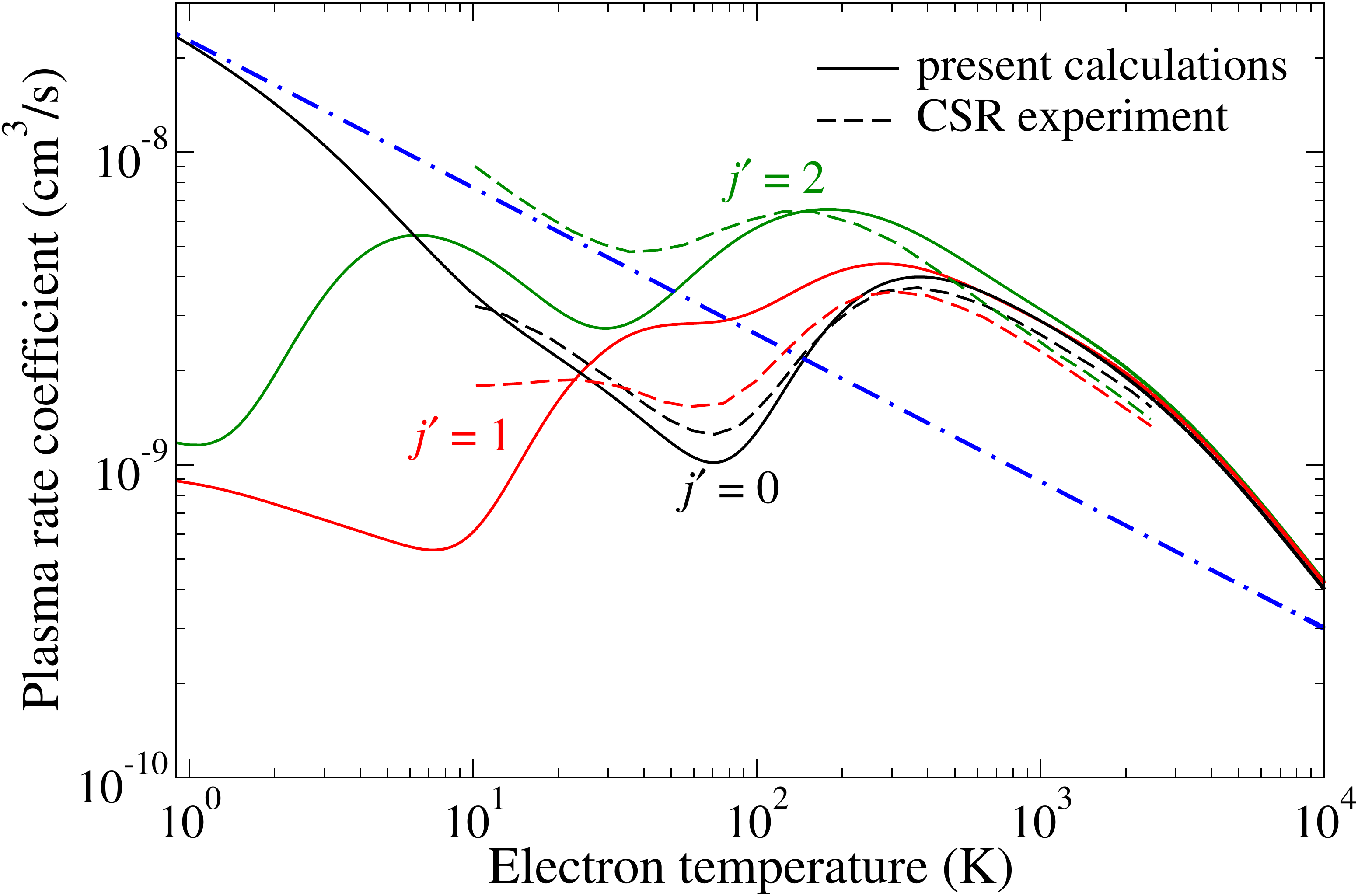}
\caption{\label{fig-plasm}
(color online). Plasma rate coefficients are shown for different initial rotational states
$j' = 0, 1, 2$. The solid curves result from the present theory, while the dashed curves
are taken from the experiment at the CSR \cite{Novotny_HeH_Science_2019}. The dot-dashed
line represents data employed by \citet{Guesten_HeH_Nature_2019} in their chemistry
kinetics simulations of NGC 7027.
}
\end{figure}

A last bit of information shown in the top panel of Fig.~\ref{fig-drj} is the decomposition 
of the computed DR rate into channels with different total angular momentum $J$ in 
a manner similar to Eq.~(\ref{eq-sigma-sumJ}).
However, all the angular factors are included in the presented data, so the red curve
is just a direct sum of the $J=0,1,2$ curves. It is clear that the peak at 45 meV is
created dominantly in the $J=1$ channel which is in agreement with the experimentally 
observed angular distribution of fragments \cite{Novotny_HeH_Science_2019} formed
by the rotational $j=1$ angular shape.

Plasma rate coefficients are obtained by an average of the state-specific DR rate
over the Maxwellian distribution of the colliding electrons. The computed plasma rates
are shown in Fig.~\ref{fig-plasm} for the lowest initial rotational states $j'=0, 1, 2$.
The $j'=0$ results compare very well with the plasma rate coefficients derived from the CSR
experiment \cite{Novotny_HeH_Science_2019}. The discrepancies for $j'=1, 2$ visible below
100 K can be linked with the discrepancies seen in Fig.~\ref{fig-drj} below 10 meV.

The dot-dashed line represents the temperature dependence of the plasma rate coefficient
employed in the temperature and density simulations of NGC 7027 \cite{Guesten_HeH_Nature_2019}.
These simulations were carried out to estimate the
emissivity of the $j = 1 \rightarrow 0$  line of HeH$^+$ ions (having the functional form: 
$k_2 = 3.0 \times 10^{-10} (T/10^4)^{-0.47}$ cm$^3$/s). Our present calculations estimate
the plasma rate coefficient at 10$^4$ K to be slightly above $4.0 \times 10^{-10}$ cm$^3$/s.
These results therefore suggest that the unexpectedly high brightness of the
$j = 1 \rightarrow 0$  transition observed in the first HeH$^+$ detection is not caused by the
slower destruction of the HeH$^+$ cations via the DR process.

We thank Old\v{r}ich Novotn\'{y} and Andreas Wolf for helpful discussions and for communicating data prior to publication.  
The work of CHG has been supported by the U.S. Department of Energy, Office of Science, 
under Award No. DE-SC0010545; Basic Energy Sciences. R\v{C} and DH acknowledge support of 
the Grant Agency of Czech Republic (Grant No. GACR 18-02098S).

\bibliographystyle{apsrev}
\bibliography{DR}
\end{document}